# Quantum Mechanics of Human Perception, Behaviour and Decision-Making: A Do-It-Yourself Model Kit for Modelling Optical Illusions and Opinion Formation in Social Networks

Ivan S. Maksymov

Artificial Intelligence and Cyber Futures Institute, Charles Sturt University, Bathurst, NSW 2795, Australia

E-mail: imaksymov@csu.edu.au

> *"Madness is something rare in individuals — but in groups, parties, peoples, epochs it is the rule"*
> *Friedrich Nietzsche*

**Abstract**: On the surface, behavioural science and physics seem to be two disparate fields of research. However, a closer examination of problems solved by them reveals that they are uniquely related to one another. Exemplified by the theories of quantum mind, cognition and decision-making, this unique relationship serves as the topic of this chapter. Surveying the current academic journal papers and scholarly monographs, we present an alternative vision of the role of quantum mechanics in the modern studies of human perception, behaviour and decision-making. To that end, we mostly aim to answer the '*how*' question, deliberately avoiding complex mathematical concepts but developing a technically simple computational code that the readers can modify to design their own quantum-inspired models. We also present several practical examples of the application of the computation code and outline several plausible scenarios, where quantum models based on the proposed do-it-yourself model kit can help understand the differences between the behaviour of individuals and social groups.

**Introduction**

In the late 19th century, many scientists believed that physics had achieved almost everything it could give to humankind. However, a fundamental error produced by a classical physical theory challenged that point of view, heralding the advent of quantum mechanics (Kragh, 2000). In the 20th century, quantum mechanics served as a key tool for scientists from different disciplines to discover new physical effects and create novel devices. Nowadays, we are witnessing the advent of commercial quantum technologies that revolutionise the way we live and work, enabling new functionalities that extend well beyond the abilities of any existing computer (Reinhold & Friis, 2023).

Yet, intriguingly, quantum mechanics has become an innovative emerging tool used by psychologists and behavioural scientists to understand and predict how humans form opinions, perceive the world and make decisions (Mindell, 2012; Busemeyer & Bruza, 2012). This chapter introduces the readers to the fascinating world that entangles behavioural science, psychology and physics, demonstrating the advantages of quantum models of human perception, behaviour and decision-making over the traditional theoretical approaches. The discussion also demonstrates that the principles of quantum mechanics can be used to develop games and virtual reality systems that teach students, including astronauts, pilots and operators of unmanned aerial vehicles, about human and computer vision systems, object detection and identification, visual field construction, autonomous movement and strategy (Khrennikov, 2006; Yamamoto & Yamamoto, 2006; Atmanspacher & Filk, 2010; Busemeyer & Bruza, 2012; Aerts & Arguëlles, 2022; Maksymov, 2024).

The text of this chapter is organised differently from any previous relevant works that mostly adopt one of the following pivotal points of view—psychological, mathematical, computational or data science-centred. Instead, it discusses the fundamental aspects of quantum mechanics and reveals their connection with essential patterns of human behaviour. While some knowledge of physics and mathematics is required to understand the forthcoming discussion, we will endeavour to use the language adopted in high-school science curricula. We will also refer to a computational code that the readers may freely access, run and modify to produce their own quantum-mechanical model of human behaviour.

**Quantum mechanics and human behaviour**

Mechanics is a branch of classical physics that studies motion caused by forces applied to physical objects such as balls, cars and celestial bodies. Quantum mechanics describes the motion of objects at the atomic and subatomic scales (Feynman, 1964). Also known as quantum theory, quantum mechanics can describe many systems that classical physics cannot, including photons and electrons. It also enables many technologies that underpin our modern lifestyles, ranging from semiconductor chips and medical imaging to optical fibre communication systems and quantum computers (Nielsen & Chuang, 2002; Reinhold & Friis, 2023).

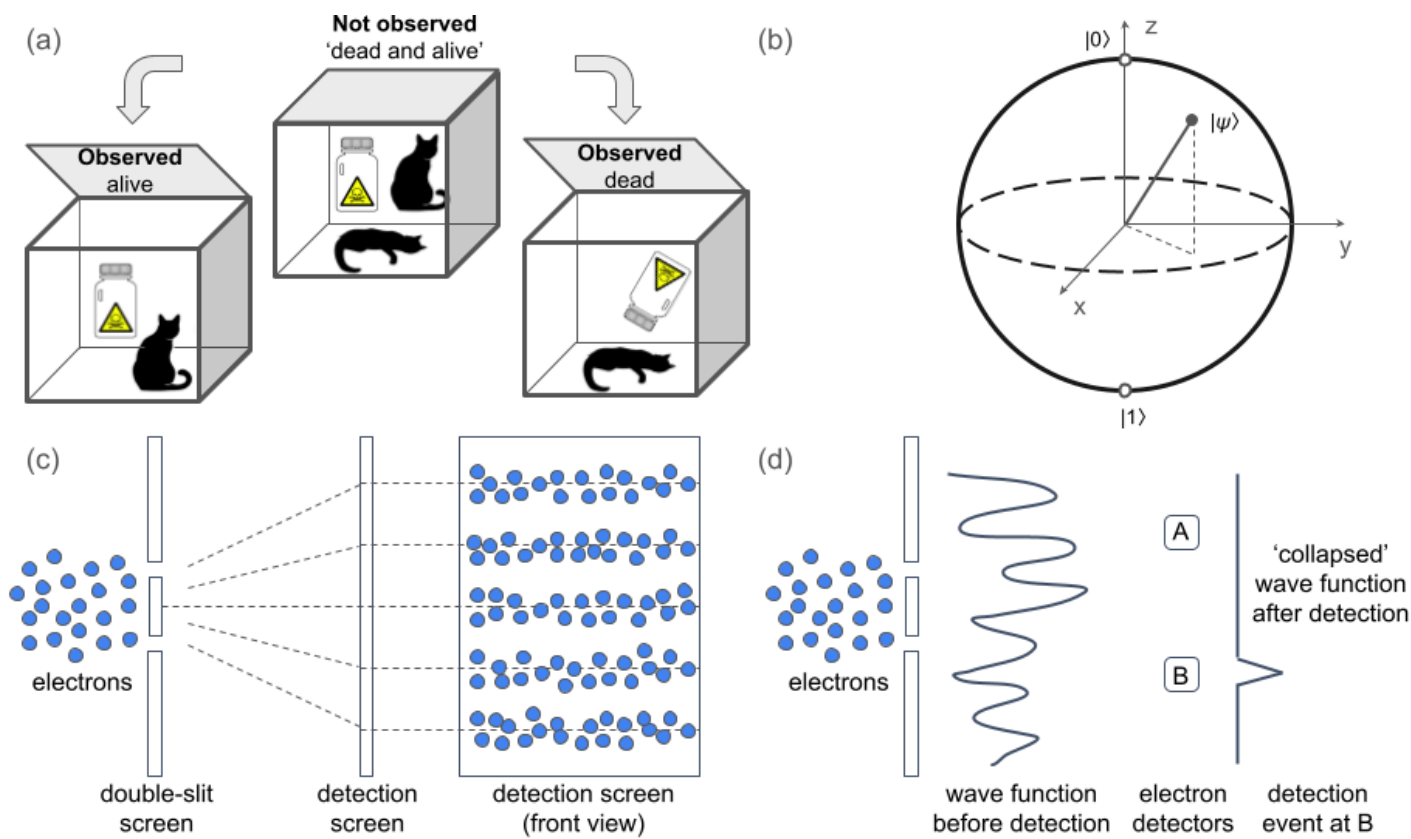

***FIG. 1:*** *(a) Schrödinger's cat thought experiment. If you place a cat and a bottle with poison in an opaque box and seal it, you cannot know whether the cat is dead or alive until you open the box. In other words, until the box is opened, the cat is 'dead and alive' at the same time. (b) Illustration of a projective measurement of a qubit $|\psi\rangle$ using the Bloch sphere. (c) The double slit experiment that demonstrates that the electrons can interfere with each other as they pass through the slits and spread out in waves. (d) Illustration of the collapse of the wavefunction as a result of detection.*

Quantum mechanics is difficult to understand using the established principles of classical physics (Feynman, 1964). One of the major conundrums that it offers is the concept of superposition (Schilpp, 1949; Feynman, 1964), which is the ability of a particle to exist in a range of possible states, famously exemplified by Schrödinger's cat (Figure 1a). Heisenberg's uncertainty principle—a fundamental concept in quantum mechanics—is another example. It states that there is a limit to the precision with which certain pairs of physical properties, such as position and momentum, can be simultaneously measured (Feynman, 1964). This means that in an experiment aimed to measure one property more precisely the other property can be known only with lower precision. To give a bigger picture, we should also mention quantum entanglement, which is the phenomenon that occurs when a pair of particles is created such that the quantum state of each individual particle cannot be described independently of the state of the other, including when the particles are separated by a large distance.

While our further discussion will clarify the physical meaning of Schrödinger's cat experiment and other fundamental principles, we are already well-positioned to intuitively say that quantum mechanics and human behaviour have something in common. Indeed, the same person can behave differently in different situations, which means that the behaviour cannot be observed as a whole but only certain characteristics of it become apparent in specific circumstances (Patry, 2011). Thus, we may assume that there exists a range of possible human mental states that may be used to describe the behaviour as a whole. However, in a specific situation we can observe only certain characteristics of the behaviour. Yet, if we design an experiment so that it accurately highlights one particular behavioural characteristic, the other characteristics will be measured less accurately or they may not even be visible at all (Fleeson & Wilt, 2010).

The influence of the environment on behaviour has also been a subject of philosophical studies, including the famous maxim pronounced by Ortega y Gasset: '*I am me and my circumstance*' (Ortega y Gasset, 1966). Further connections between Ortega y Gasset's circumstance and quantum aspects of cognition have been established in more recent works (Castro, 2013). Generally speaking, the representation of the whole behaviour as a quantum system is also consistent with both notion of discrete mental states (Aerts & Arguëlles, 2022; Aerts and Beltran, 2022) and understanding of information as energy states of a physical system (Wilson, 2000; Dittrich, 2014). As demonstrated below, these physical principles hold the potential to describe the differences in behaviour of individuals and social groups, opening up opportunities to address long-standing philosophical questions relating to morals, truth and values in the contexts of societies and cultures (Nietzsche, 1886).

Previous research demonstrated that situation-specific characteristics of behaviour can be understood using methods of classical physics (Galam, 2012). While classical models have been successfully applied in practice (Galam, 2012), further theoretical attempts to improve them created a family of hypotheses proposing that physical laws and interactions from the realm of classical mechanics alone cannot fully explain human mind and consciousness, suggesting instead that quantum-mechanical phenomena should serve as a more accurate model of behaviour (Khrennikov, 2006; Busemeyer & Bruza, 2012; Mindell, 2012; Wendt, 2015; Pothos & Busemeyer, 2022). While the quantum mind theory mostly belongs to the realms of philosophy and mathematics, the attempts to apply quantum mechanics in the studies of optical illusions (Atmanspacher & Filk, 2010; Busemeyer, & Bruza, 2012), decision-making paradoxes (Aerts, Sandro & Tapia, 2012; Aerts, Sandro & Tapia, 2014; Martínez-Martínez, 2016; Wei, al Nowaihi & Dhami, 2019; Moreira, 2020) as well as social and political behaviour (Tesař, 2015; Allan, 2018) have demonstrated that quantum-mechanical models not only successfully reproduce the previous results obtained using the classical models but also offer additional degrees of freedom for scientists to describe human behaviour with a higher accuracy.

The demonstrations of the agreement between the predictions of the quantum-mechanical and classical physical models of human behaviour correlates with the cornerstone principles of a mainstream quantum theory called Quantum Darwinism that aims to explain the emergence of the classical world from the quantum world following a process of Darwinian natural selection (Zurek, 2003; Zurek, 2009). In quantum mechanics, decoherence refers to an alternation of the state of a quantum system caused by the system's interaction with the environment. As a result of this process, the system becomes entangled with its environment in some unknown way and its description becomes impossible without describing the state of the environment. Quantum Darwinism theory proposes that decoherence is caused by the interaction of the system with its environment and not by the observation of it. For example, according to the theory, the environmental influences explain why we do not observe big objects being in quantum states. While any further discussion of pros and cons of Quantum Darwinism (Zurek, 2015) would lie beyond the scope of this text, this theory may explain to us why under specific circumstances we can observe only some characteristics of the behaviour and why it suffices to use the methods of classical physics to understand these particular characteristics with acceptable accuracy. However, if we want to understand the behaviour as a whole, we should use the principles of quantum mechanics because the whole behaviour originates from a superposition of quantum-like mental states and, hence, can be explained only using the methods of quantum mechanics.

To put this discussion into further perspective, let us compare the operation of a traditional digital computer with the operation of a quantum computer (Nielsen & Chuang, 2002). Similarly to an on/off light switch, a bit of a digital computer is always in one of two physical states corresponding to the binary values '0' and '1'. However, a quantum computer uses a quantum bit (qubit) that can be in states $|0\rangle$ and $|1\rangle$. While these states are analogous to the '0' and '1' binary states, a qubit also exists in a continuum of states between $|0\rangle$ and $|1\rangle$, i.e. its states are a superposition $|\psi\rangle = \alpha|0\rangle + \beta|1\rangle$ with $|\alpha|^2 + |\beta|^2 = 1$.

From the physical point of view, the states of the qubit can be illustrated using the Bloch sphere (Figure 1b). When a closed qubit system interacts in a controlled way with the environment, the experiment reveals that the measurement probabilities for $|\psi\rangle = \alpha|0\rangle + \beta|1\rangle$ are $P_{|0\rangle} = |\alpha|^2$ and $P_{|1\rangle} = |\beta|^2$, which means that the qubit is in one of its basis states. Graphically, the measurement procedure means that the qubit is projected on one of the coordinate axes (e.g., z-axis in Figure 1b). The same picture can be used to illustrate the relationship between behaviour and the principles of quantum information processing.

**Physics-driven model of quantum human behaviour**

The interference fringes observed in a famous double-slit physical experiment provide another example of the superposition principle (Walker, 2021). Therefore, after a general explanation of the physics that underpins this experiment, we will use the double-slit configuration as a toy model of human behaviour.

*Double-slit experiment*

Atoms are the basic building block of matter. However, towards the end of the 19th century the concept of atoms was rather controversial (Smith, 2020). In an attempt to test the existence of atoms, the German physicist Max Planck studied the properties and behaviour of blackbody radiation (a blackbody is an object that absorbs all radiation falling on it). Planck concluded that a blackbody absorbs and re-emits radiation in hypothetical discrete bits that he called quanta (Kragh, 2000).

In 1905, Albert Einstein provided further arguments speaking in favour of the existence of the quanta, suggesting that radiation itself comes in discrete bits of energy called the photons. Einstein's hypothesis posed a problem since there was a well-established body of experimental evidence speaking in favour of a wave theory of light. Then, in 1923, Louis de Broglie demonstrated a direct mathematical relationship between an electron's wave-like property (wavelength) and a particle-like property (momentum). Later, in 1926, Erwin Schrödinger used the classical theory of waves and some quantum conditions from de Broglie's relationship. The result was the Schrödinger wave equation, where the motion of a particle such as an electron is calculated from its wave function.

The wave function is a mathematical description of the quantum state of an isolated quantum system. In classical mechanics, we can readily interpret the concepts represented in the theory (the physical observables such as energy and momentum) and their relation to the objects that possess them (Feynman, 1964). However, the interpretation of the concepts of quantum mechanics requires us to perform a specific mathematical operation on the quantum state of the electron. Such operations can be regarded as mathematical algorithms that enable us to know the wave function by releasing the observable.

To demonstrate that light is a wave, scientists illuminate a narrow slit made in an opaque screen to show that the light moves through the slit, bends around at its edges and then spreads out beyond it. This process is called diffraction. When the screen has two narrow slits, the optical waves diffracted by the two slits produce an alternating pattern of light and dark bands called interference fringes. These experimental results can be reproduced using the waves of different nature, including water waves (Feynman, 1964).

The same experiment can be conducted using electrons (Figure 1c). Each electron passing through the slits is registered on the screen as a single bright spot. As more and more electrons pass through the slits, the bright individual spots start to group together, overlapping and merging. As a result, a double-slit interference pattern of alternating bright and dark fringes is formed similarly to the pattern observed in double-slit experiments involving optical waves. Thus, each individual electron behaves as a wave, described by a wave function $\psi$, that passes through both slits simultaneously and interferes with itself before striking the screen.

The square magnitude of the wave function $|\psi|^2$ represents the probability density of the particle. Thus, the alternating peaks and troughs of the electron's wave function translate into a pattern of quantum probabilities: a bright fringe corresponds to a higher probability of finding the next electron and a dark fringe corresponds to a lower probability. Before an electron strikes the screen, it has a probability of being found 'anywhere' where the modulus square of the wave function is bigger than zero. This probability of many states existing at the same time results in quantum superposition.

Now we return to the analogy between quantum mechanics and human behaviour. Both observation of a particular characteristic of human behaviour under certain circumstances and analysis of human behaviour considered as a whole can be modelled similarly to the process illustrated in Figure 1d. The state of the electron is not defined until the wave function interacts with the screen, at which point it 'collapses' and the electron appears in only one place. We can say the same about the behaviour: it is not defined until it is measured under certain conditions. Thus, the double-slit structure can be considered as the environment with which the individual interacts but the electron detector represents the circumstance that shapes the behaviour. Several specific examples of such a modelling are presented below.

*The model*

We numerically solve the Schrödinger equation in two-dimensional space. This equation is a linear partial differential equation that governs the wave function of a quantum-mechanical system. Its understanding requires knowing the concepts and notations of calculus, including derivatives with respect to space and time. However, in the following we will avoid discussing such mathematical concepts. Instead, we will employ the Crank-Nicolson method, a well-known computational algorithm for solving certain classes of partial differential equations (Thomas, 1995). An Octave/MATLAB code that implements this algorithm can be accessed following the instructions given at the end of this text. Additional technical details can be readily found in the literature (Thomas, 1995) and on dedicated open source software development platforms (Mena, 2023). It is also noteworthy that the results discussed below can be obtained using similar numerical techniques such as the finite-difference time-domain method (Sullivan, 2000; Larsen, 2022).

In brief, we use a two-dimensional spatial grid that consists of $N$ points along the $x$-coordinate direction and $N$ points along the $y$-coordinate direction. We also consider $N_t$ time points. Subsequently, the continuous wave function $\psi$ describing the system at a given spatial point and instant in time will be an approximate discrete function $\psi^k_{i,j}$, where $i$ and $j$ are the spatial indices and $k$ is the index of the instance of time. Using the two-dimensional mesh and the discretisation in time, we approximate the time and spatial derivatives of the Schrödinger equation employing the Euler's method (Thomas, 1995). After some algebraic derivations, we obtain a matrix equation that can be solved using the standard computational procedures available in scientific and technical programming languages.

MATLAB is a programming language that manipulates matrices to implement algorithms, process data and plot functions (Octave is a free version of MATLAB). In our algorithm, we represent an electron as an energy wave packet that has a two-dimensional Gaussian shape (also known as bell-shaped function). The double-slit structure is represented as a potential barrier for the electron. The height of the barrier is a model parameter that can be tuned to create customised simulation scenarios. The barrier can also be removed from the model by setting its height to zero. The boundaries of the two-dimensional space are the walls of an infinitely high potential barrier, which means that the electron cannot leave the computational domain.

As the first test, we use the model to reproduce the result predicted in Figure 1c. Plotting the probability density $|\psi|^2$ in the two-dimensional space, we can see a Gaussian wave packet that first moves from the left side of the computational domain towards the double-slit structure and then interacts with the slits, forming a pattern of interference fringes near the right wall (Supplementary Video 1). The readers can repeat the simulation for different values of the height of the potential barriers that form the double-slit structure. The execution of the code with different barrier parameters will reveal that the pattern of interference fringes changes as a function of the barrier height (in some situations, the wave packet can be fully reflected off the barrier). A similar result can be obtained for the potential barriers of different widths. The readers may also either remove one slit or add more slits to obtain a more complex interference picture.

As the second test, we remove the double-slit structure by setting the height of the potential barriers to zero. In that simulation, the wave packet moves from left to right, slightly spreading in space, until it reaches the right boundary (wall) and reflects from it (Supplementary Video 2). The reader will observe an interesting and fundamental physical result: the solution of the Schrödinger equation that describes the time evolution of a wave packet bouncing off the wall illustrates interference of the reflected part of the packet with its incident part, resulting in rapid oscillations in the probability density near the wall (Andrews, 1998).

How can these physical results be interpreted in the framework of the quantum mind? The wave packet represents a single particle that models the behaviour of an individual. The corresponding wave function can be called the mental wave function but the observed physical interference process can be called the interference of mind (Khrennikov, 2006). Thus, in agreement with the previous quantum mind hypotheses, our model demonstrates that the mind of a single individual can interfere with itself. This result also correlates with the mainstream theories developed in the fields of psychology (Bompas, 2020; Qureshi, 2020) and philosophy (Dicker, 1993). Moreover, the model enables us to simulate the behaviour of an individual who interacts with the environment that can be represented by a simple wall or an arbitrary complex structure (called potential wells) that represents a real-life situation, an example of which will be given below.

It is straightforward to modify the computational code to add another individual that can be modelled using either an identical two-dimensional Gaussian wave packet or wave packet with different parameters. That is, by changing the parameters we can model an individual with a different mental wavefunction. For the sake of illustration, let us assume that the wave packet corresponding to the second individual moves from right to left such that it should collide with the first wave packet in the middle of the computational domain. By running this simulation, we will observe the interference of the two wave packets (Supplementary Video 3). The interaction of three or more wave packets can be modelled similarly to demonstrate the interference of many minds. Finally, this simulation scenario can be extended by adding a double-slit structure or any other potential barrier, the presence of which will influence the interference process. Such a complex simulation scenario will take into account both interaction between individuals and influence of the environment on the individuals.

# Examples of application of the quantum model

*Optical illusions*

Optical illusions have always fascinated scientists and members of the general public (Necker, 1832; Washburn, Reagan & Thurston, 1934; Shapiro & Todorović, 2017). Different types of optical illusions also inspired artists, photographers and advertisers (Fisher, 1967; Lindstrøm & Kristoffersen, 2001). In particular, such ambiguous images as the Necker cube (Necker, 1832), Rubin's vase (Parkkonen, Andersson, Hämäläinen & Hari, 2008; Khalil, 2021), 'My Wife and My Mother-in-Law' (Nicholls, Churches & Loetscher, 2018), Rabbit-Duck illusion (McManus, Freegard, Moore & Rawles, 2010) and Spinning Dancer illusion (Troje & McAdam, 2010) have been a paradigmatic topic of scientific research in the fields of psychology (Long & Toppino, 2004; Kornmeier & Bach, 2005; Busemeyer & Bruza, 2012; Stonkute, Braun, & Pastukhov, 2012; Meilikhov & Farzetdinova, 2019) and psychiatry (Basar-Eroglu, Mathes, Khalaidovski, Brand, & Schmiedt-Fehr, 2016). The other areas where optical illusions are of interest to scientists and engineers is the development of computer vision and robotic systems capable of perceiving the world like a human (Takeno, 2013). A realistic model of optical illusions can also be used to improve video games and virtual reality systems (Wang, Xu, Wang, Huang, Chang, Cheng, Lin, & Cheng, 2021; Maksymov & Pogrebna, 2023; Magnuski, 2023). Yet, optical illusions play an important role in the training of pilots and astronauts (Yamamoto & Yamamoto, 2006; Clément, Allaway, Demel, Golemis, Kindrat, Melinyshyn, Merali, & Thirsk, 2015).

Recently, optical illusions have attracted the attention of researchers working on quantum models of cognition, perception and decision-making (Atmanspacher & Filk, 2010; Busemeyer, & Bruza, 2012; Pothos & Busemeyer, 2022; Kauffman & Roli, 2022). One of the key optical illusions studied in the cited works is the Necker cube (Figure 2a, left), an ambiguous figure that has served as a paradigmatic example in psychology (Kornmeier & Bach, 2005) and helped demonstrate that the human brain operates as a neural network with two distinct equally possible interchangeable stable states (Inoue & Nakamoto, 1994; Ward & Scholl, 2015).

In a typical experiment involving a Necker cube, the observer is asked the question 'Is the shaded face of the cube at the front or at the rear?'. Using an electric button or any other device that can register a binary 'front-rear' response, the observer's typical answers result in a series of interpretations that randomly switch between 'front' and 'rear' (in the following, we will denote these states as $|0\rangle$ and $|1\rangle$, respectively). When the result reported by the observer is plotted as a function of time (Figure 2a, right, the dashed line), one obtains a signal consisting of rectangular pulses of random duration. The duration of the pulses reported by different observers depends on the observer's age, gender and other factors (Lo & Dinov, 2011). However, the general square pulse pattern shown on the right in Figure 2a remains virtually the same in many experiments. Human perception of the other known optical illusions is investigated employing the same experimental technique, also resulting in a series of square-like pulses of random duration (Ward & Scholl, 2015; Wang, Sang, Hao, Zhang, Bi & Qiu, 2017).

Intriguingly, when the electrical brain activity and eye movement are recorded simultaneously with the vision of optical illusions reported by the observer (Piantoni, Romeijn, Gomez-Herrero, Van Der Werf, & Van Someren, 2017; Joos, Giersch, Hecker, Schipp, Heinrich, van Elst, & Kornmeier, 2020), the combined result that integrates data from all three measurements suggests that the observer's perception may be in a superposition of the states $|0\rangle$ and $|1\rangle$ (Figure 2a, right, the solid line). That is, while the observer can report only one of the two binary states of the Necker cube, the electroencephalogram and eye-tracking data indicate that the observer might simultaneously see the two states of the cube. This situation resembles the hypothetical Schrödinger's cat experiment, where an attempt to verify the state of the cat results in the observation of one of the two possible states – 'dead' or 'alive'. In the case of the Necker cube, the verification is done by the observer who presses a button to report whether they see the cube in the state |0⟩ or |1⟩.

A Schrödinger equation-based quantum model of superposition of the perception states of the Necker cube was proposed by Busemeyer & Bruza (2012). In contrast to a classical Markov model of bistable perception of the Necker cube where the possible states of the cube could accept only the binary values |0⟩ and |1⟩, the quantum model demonstrated harmonic oscillations of the perception between the two binary states. Importantly, in agreement with the experimental brain activity and eye-tracking data, the predicted oscillatory behaviour implied that, at some moments of time, the perception could be in a superposition of the states |0⟩ and |1⟩. Conceptually the same result was also obtained independently using a quantum 'Necker-Zeno' model (Atmanspacher & Filk, 2010).

Harmonic oscillators are ubiquitous in physics and many realisations of such oscillators can be found in mechanical systems, electrical circuits and electromagnetic wave phenomena. A simple harmonic oscillator is a particle that undergoes harmonic motion about an equilibrium position, such as a marble rolling in the bottom of a bowl (Figure 2b, left). Mathematically, in a one-dimensional system, such a motion can be studied considering the potential energy of the oscillator to be a parabolic function of its position.

However, the classical formulation of the harmonic oscillator cannot be used to describe physical systems where quantum effects play an important role. Subsequently, one has to solve the Schrödinger equation for a particle trapped in a parabolic potential well. The solution of this equation enables us to find the allowed energies $E$ and their corresponding wave functions $\psi$. Simply citing the result of the solution, it can be demonstrated that the allowed energies are $E_n = (n + 1/2)\hbar\omega$, where $n = 0, 1, 2, 3$ and so on, $\hbar$ is Plank's constant and $\omega$ is the angular frequency of the oscillator. Several fundamentally important features appear in this solution. For example, unlike a classical oscillator, the energies of a quantum oscillator can have only evenly spaced discrete energy values.

The physics of a quantum harmonic oscillator was employed by Maksymov & Pogrebna (2023) to extend the Schrödinger equation-based quantum model of perception of the Necker cube proposed in Busemeyer & Bruza (2012). In classical mechanics, a marble that rolls back and forth inside an empty bowl cannot surmount a substantial barrier placed on its way. However, in the realm of quantum mechanics, an electron trapped in a parabolic well can pass (tunnel) through a barrier (Figure 2b, right). This physical property was employed to capture the perception dynamics of the Necker cube inferred from the experimental electrical brain activity and eye movement data.

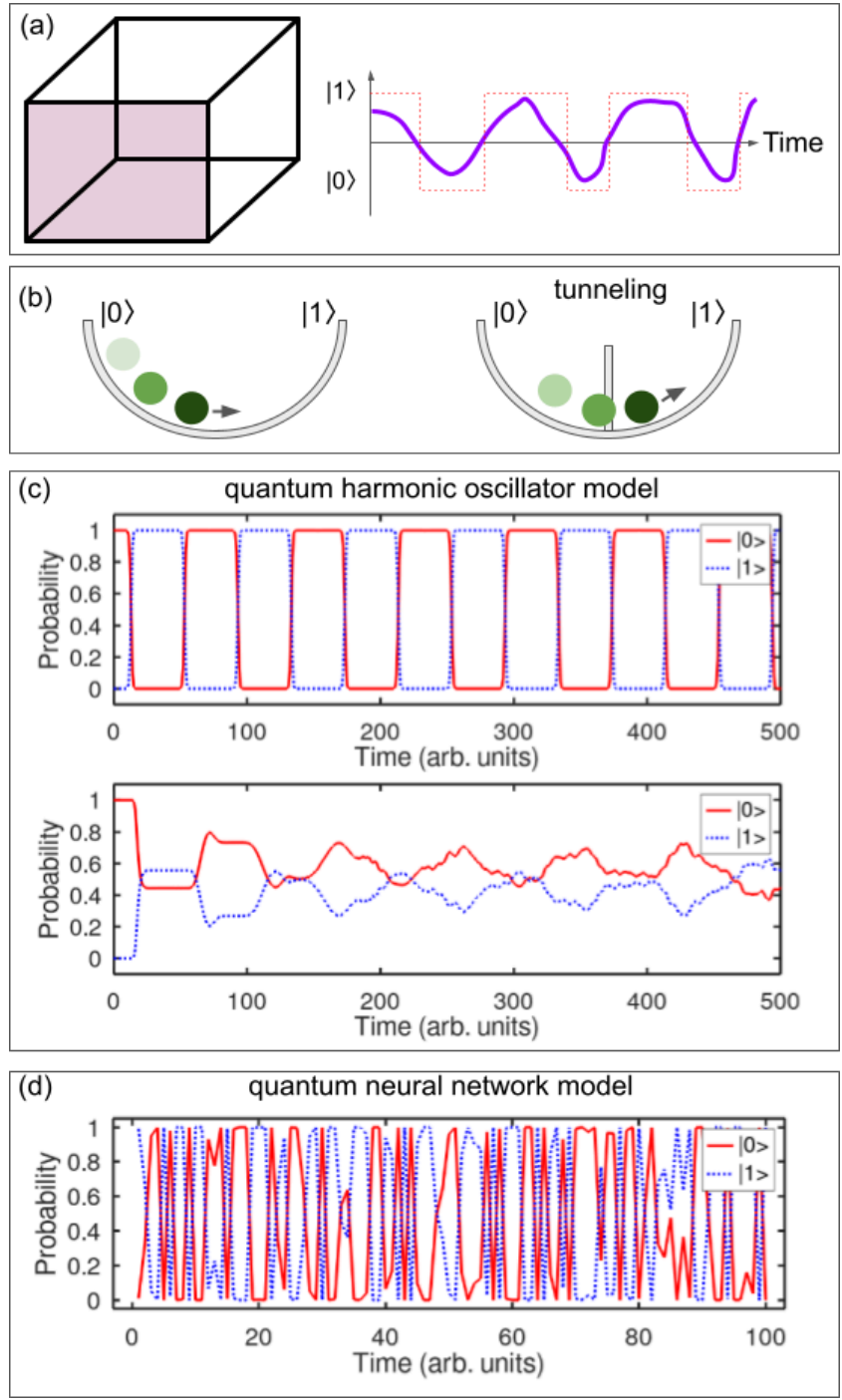

**FIG. 2:** *(a)* Left: The Necker cube and its two possible stable interpretations denoted as the states $|0\rangle$ and $|1\rangle$. Right: Schematic of a typical binary perception pattern (the dashed line), where the cube can be either in $|0\rangle$ or $|1\rangle$ state. The solid line sketches a possible continuous curve that corresponds to a superposition of states $|0\rangle$ or $|1\rangle$. *(b)* An electron trapped in a parabolic well behaves as a quantum harmonic oscillator and it can pass through a barrier due to the quantum tunnelling effect. The labels $|0\rangle$ and $|1\rangle$ denote the fundamental perceptual states of the Necker cube used in the harmonic oscillator model. *(c)* Switching of the Necker cube perception predicted by the quantum harmonic oscillator models with (top panel) parabolic potential well and (bottom panel) parabolic potential well with a barrier. *(d)* Prediction of the Necker cube perception made by the quantum neural network. Note that the time units used in panels (c) and (d) are different due to the differences in the algorithms of the quantum oscillator model and quantum neural network digital twin.

Figure 2c shows the results of the simulations conducted using a parabolic potential well (the top panel) and a parabolic well with a barrier (the bottom panel). In both simulations, the parabolic potential well was virtually split into two equal spatial regions denoted as $|0\rangle$ and $|1\rangle$. The probabilities of finding the electron inside those regions were calculated and interpreted as the probabilities of perceiving the Necker cube in the states $|0\rangle$ and $|1\rangle$, respectively. We can see that the model with the parabolic well effectively reproduced the result obtained with a more simple quantum model proposed in Busemeyer & Bruza (2012): the switching between the $|0\rangle$ and $|1\rangle$ states alternates with the moments of time when the perception is in a superposition of the states $|0\rangle$ and $|1\rangle$ [effectively, the model also reproduces the results obtained by Atmanspacher & Filk (2010)]. Moreover, the addition of a barrier to the parabolic potential well and the introduction of the quantum tunnelling effects in the model enabled more accurate modelling of the perception with a better agreement with the predictions made based on the experimental brain activity and eye-tracking data.

The result in the bottom panel of Figure 2c also agrees with the conclusions made in the previous work (Wilson, Hecker, Joos, Aertsen, Tebartz van Elst, & Kornmeier, 2023), where it has been shown that perception can become unstable before the actual reversal event is reported by the observer. Experiments also demonstrated that observers may unconsciously control their perception of the cube, thus changing the number of perceptual reversals over time (Long & Toppino, 2004). However, it is known that the reversal cannot be prevented entirely (Long & Toppino, 2004). This fact is also reproduced by the model.

The results produced by the quantum oscillator model were confirmed using a deep neural network digital twin (Maksymov, 2024). The neural network consisted of an input layer with 100 input nodes, three hidden layers each with 20 nodes each and an output layer with two output nodes (Kim, 2017). The input nodes encoded an image of the Necker cube but the two output nodes were used to classify the perceptual states $|0\rangle$ or $|1\rangle$ of the cube.

While the physical processes underpinning the dynamics of switching between the perceptual states of the Necker cube remains a subject of debate (Kornmeier & Bach, 2005), one of the currently accepted theories suggests that the switching can be explained by chaotic processes observed in nonlinear dynamical systems (Inoue & Nakamoto, 1994; Shimaoka, Kitajo, Kaneko & Yamaguchi, 2010; Chen, Xiong, Zhuge, Li, Chen, He, Wu, Hou & Zhou, 2023). The fact the brain is a dynamical system that exhibits a complex nonlinear and chaotic behaviour at multiple levels (McKenna, McMullen & Shlesinger, 1994; Korn & Faure, 2003) also speaks in favour of these theories.

To introduce a chaotic dynamical behaviour in the neural network model, a quantum-physical generator of true random numbers (Symul, Assad & Lam, 2011) was used to define the connections of the neural network. In contrast to pseudo-random number generators that are often encountered in standard scientific computation software, true random numbers are generated in real-time in a lab by measuring the quantum fluctuations of the vacuum. While in classical physics a vacuum is considered to be a space that is empty of matter, from the point of view of quantum physics the electromagnetic field of the vacuum exhibits random fluctuations in phase and amplitude at all frequencies. These fluctuations are measured and converted into random numbers that are then broadcast on the Internet and can be accessed by the users. Alternatively, true random numbers can be generated by a supercomputer (Kumar & Pravinkumar, 2023). Independently of the origin of true random numbers, a neural network based on the quantum random number generation is not biased towards one of the possible perceptual states of the Necker cube but its predictions do not repeat in time (Maksymov, 2024).

As shown in Figure 2d, the quantum neural network model predicts a chaotic pattern of switching between $|0\rangle$ and $|1\rangle$ states of the Necker cube, revealing that the perception can be in a superposition of these states. This result demonstrates that both neural network model and quantum oscillator model produce a qualitatively the same result. Indeed, a careful inspection of the switching pattern predicted by the neural network reveals that it combines the two patterns produced by the quantum oscillator models with and without the potential barrier.

For example, while in the time range from $T = 0$ to $T = 30$ the forecast of the neural network resembles the quasi-periodic output of the quantum oscillator model with a parabolic potential well, the pattern predicted by the network at $30 < T < 60$ is in good agreement with the output of the quantum oscillator model with the potential well with a barrier. It is noteworthy that the time units used in the quantum oscillator model and neural network are different, which means that the timescale of alternations between the perceptual states is different. However, this discrepancy is inconsequential for the current discussion since it can be eliminated using a different profile of the parabolic potential wells.

The current research conducted in this direction is focused on achieving a better agreement between the predictions of the neural network and the outcomes of quantum oscillator modelling. In particular, it has been suggested that the traditional deep learning network architectures may be replaced by a reservoir computing network that exploits a biological brain-inspired nonlinear dynamics system (Maksymov, 2024a). Both quantum oscillator model and neural network digital twin also require further improvement aimed to achieve a better agreement with experimental data. In the case of the neural network, such an improvement can be done by training the network on experimental data that combine observer's reports of visual perception of ambiguous figures with simultaneously captured eye-tracking and brain activity data sets, which is a task that can be accomplished by a reservoir computing network that operates according to the laws of quantum mechanics (Abbas & Maksymov, 2024). In its turn, the quantum oscillator model can be advanced by taking into account more complex geometries of potential wells. For example, the computational code that accompanies this text can be modified to create a two-dimensional potential well that has three, four or even more valleys and/or barriers. Assigning different states ($|0\rangle$, $|1\rangle$, $|2\rangle$ and so forth) to each valley and simulating the probability of finding the electron in each of them, one can create a model of multistable perception. The so-designed model can be used in the studies of more complex optical illusions, including depth-related tristable stimuli (Wallis & Ringelhan, 2013). Yet, a multivalley potential well model can be used to undertake a categorisation-decision computational task exemplified by images of four human faces classified as 'good guy' group and 'bad guy' group (Busemeyer & Bruza, 2012).

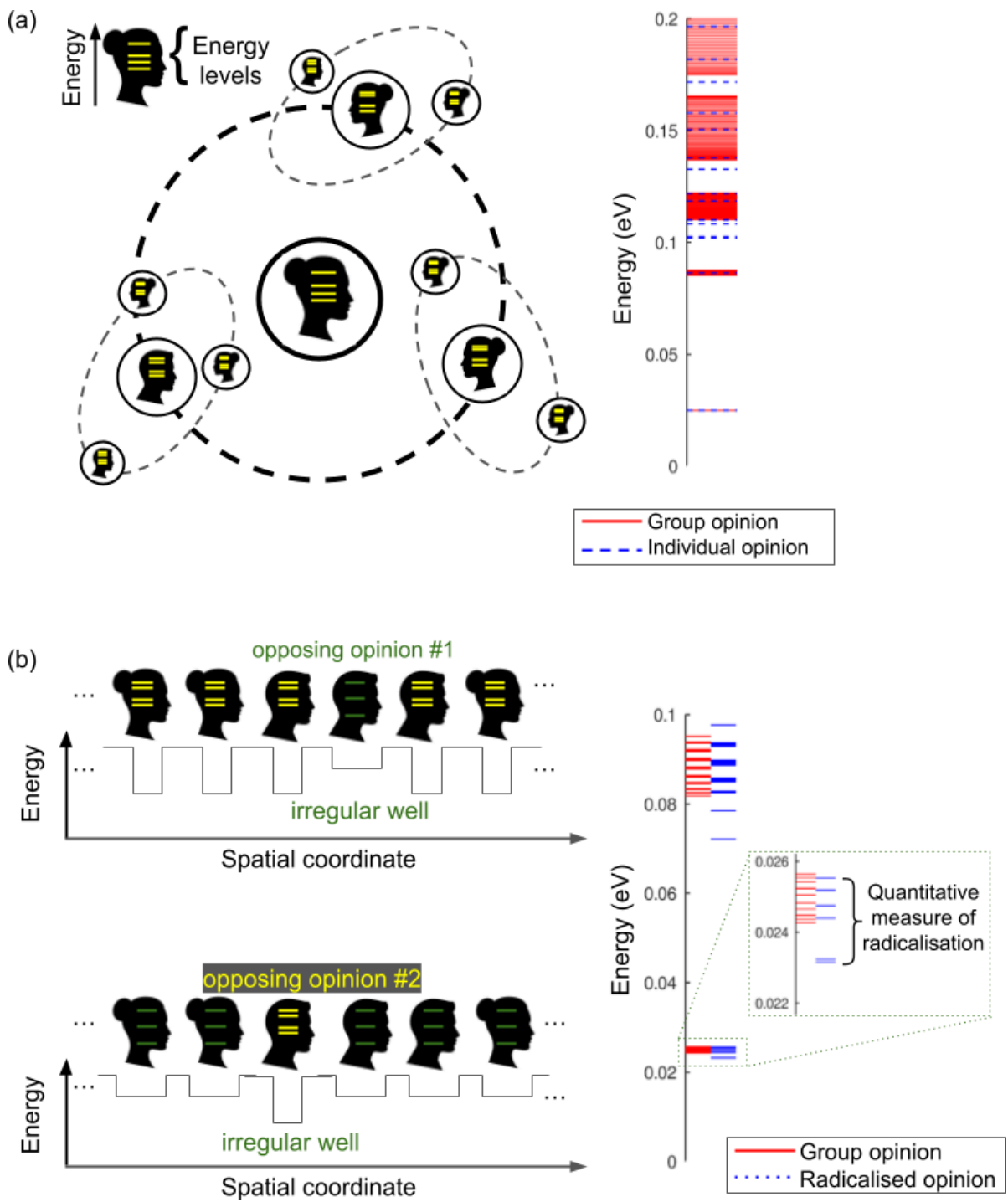

**FIG. 3:** *(a) Illustration of a social network and human systems of beliefs represented by discrete energy levels (schematically shown as the horizontal lines superposed on the human head silhouettes). The right panel shows the result of a rigorous simulation, where the energy levels corresponding to an isolated individual are denoted by the dashed lines but the energy levels of the social network are denoted by the solid lines. The grouping of the solid lines is used to model the opinion formation in the network. (b) The exposure of the social network to an opposing opinion is modelled using irregular potential wells because the energy levels in such wells are different from those in the regular wells. Different types of opposing opinions can be taken into account using irregular wells of different geometry (compare the opposing opinions #1 and #2). The introduction of an irregular well leads to a different energy level grouping, which can be interpreted as the opinion radicalisation.*

*Opinion polarisation in social networks*

In the preceding discussion, we have demonstrated that a quantum-mechanical system can have only certain allowed energy levels compared with a continuum of energy states of a classical system. Based on this property and following the previous works (Aerts & Arguëlles, 2022; Aerts & Beltran, 2022), in the paper Maksymov & Pogrebna (2024) it was suggested that a system of human beliefs can be modelled as a set of discrete energy levels of a quantum-mechanical system.

In the discussion of the quantum harmonic oscillator model represented by a parabolic potential well, the allowed energy levels were evenly spaced. However, when the potential well is rectangular, the energy states are quantised as $E_n \propto (n/L)^2$, being $n = 1, 2, \ldots$ the number of the energy state and $L$ the width of the well (Smith, 2020). Thus, varying the parameter $L$ we can tune the structure of the allowed energy levels.

How can these physical properties be used to model a social network of interacting individuals? Let us assume that in our model a single rectangular potential well represents one individual. The system of beliefs of this individual is represented by the allowed energy levels corresponding to the potential well (see the top left inset in Figure 3a, left). Then, we extend the model by adding the second individual. If the second individual has the same system of beliefs, then the potential well that represents this individual will have the same geometry and the same set of allowed energy levels. However, if the beliefs of the second individual are different, then we can choose a different potential well with its corresponding set of allowed energy levels.

Repeating the same procedure, we can add an arbitrary large number of individuals, arranging them into a social network that involves individuals with arbitrary complex beliefs and opinions (Figure 3a, left). Since the so-created social network operates using the principles of quantum mechanics, solving the Schrödinger equation we can calculate how the social network will react to the introduction of new neighbours and exposure to confronting opinions, thereby enabling us to simulate real-life social network scenarios such as disagreement, conflicts and false information (Acemoglu, Como, Fagnani & Ozdaglar, 2013). Indeed, we know that electrons occupy atomic orbitals that have discrete energy levels. When atoms interact, their atomic orbitals overlap and energy levels change. Thus, when the social network is formed by two or more individuals who share the same opinion, the interaction between the individuals will result in the grouping of the different energy levels. This process is called the energy band formation and it is illustrated in the right panel of Figure 3a, where the dashed lines correspond to the energy levels of an individual but the group opinion is denoted by the solid lines (all the energy levels presented in Figure 3 are the rigorous solutions of the Schrödinger equation).

Furthermore, if we add an individual who holds an opposite view (Figure 3b, left), the energy levels will group into different energy bands (Figure 3b, right). Analysing the differences in the energy band formation, we can understand which events lead to conflicts, how fake news is spread and why people's opinions become radicalised. For instance, as shown in the left panel of Figure 3b, in the model we can control which opposing opinion is introduced into the social network: this can be done by changing the geometry of one potential well (the so-called irregular potential well) and solving the Schrödinger equation for all potential wells in the system. An example of such a calculation is presented in the right panel of Figure 3b, where we can see that the group opinion is changed when the like-minded individuals are exposed to confronting views. It is noteworthy that Figure 3b presents just one example of irregular wells. The readers may use two and more irregular wells that have different shapes, including parabolic, triangular, quartic and their combination.

Using the irregular potential well approach, it has been demonstrated that the energy level grouping effectively reproduces the essential processes of opinion radicalisation in social networks, exemplified by the backfire effect (Maksymov & Pogrebna, 2024). The backfire effect refers to the tendency of people to give more credence to evidence that supports their pre-existing beliefs (Swire-Thompson, DeGutis & Lazer, 2020; Chen, Tsaparas, Lijffijt & Bie, 2021). This phenomenon extends to beliefs that align closely with an individual's worldview. For example, an experiment involving Republican and Democrat social network users revealed that the exposure of Republicans to the views of Democrats resulted in a significant opinion radicalisation towards conservatism (Bail, Argyle, Brown, Bumpus, Chen, Hunzaker, Lee, Mann, Merhout & Volfovsky, 2018). However, Democrats exposed to the social posts by Republicans became just slightly more liberal. The other notable examples of the backfire effect concern such pressing societal issues as vaccination (Nyhan & Reifler, 2015), climate change (Dixon, Bullock & Adams, 2019) and abortion rights (Liebertz & Bunch, 2021).

In each of the aforementioned papers, social groups exhibited the backfire effect after being subjected to an opposing view. However, while the studies carried out in the domain of politics and abortion right exhibited a 'double-sided' polarisation behaviour triggered by the backfire effect (Maksymov & Pogrebna, 2024), a 'one-sided' view was demonstrated in the domains of climate change and vaccination. This discrepancy is due to missing data for these areas and it represents a challenge for computational modelling aimed at understanding the opinion dynamics. Since the quantum model of the social network can reproduce the 'double-side effect', it is plausible that training of the model on a 'one-sided' view can help calculate the missing second opinion polarisation (Maksymov & Pogrebna, 2024).

The quantum model can also be used to extend the existing classical socio-physical models that have been successfully used to predict the results of major political elections (Galam, 2005; Castellano, Fortunato & Loreto, 2009; Hu, 2017; Redner, 2019; Galam, 2022; Interian & Rodrigues, 2023). For example, in the Sznajd model (Sznajd-Weron & Sznajd, 2000) and its variations (Redner, 2019), the opinions of individuals are represented by the states ↑ and ↓ (in classical physics such states may have the meaning of magnetic dipoles). The interaction between two neighbours changes the opinions of their respective neighbours. However, the opinions of the two particular interacting neighbours are assumed to be unchanged. While other classical approaches can resolve this problem by adding more degrees of freedom to the model (Galam, 1997), the quantum-mechanical model naturally takes into account the interaction between all neighbours in the social network (Maksymov & Pogrebna, 2024).

To test the model outlined in this section, the readers are invited to modify the computational code that accompanies this text by adding rectangular potential wells and calculating the wavefunctions. Alternatively, the readers may refer to the computational code supplied in the paper Maksymov & Pogrebna (2024). Although the cited paper uses a different numerical method, the same computational principles can be applied to the computational code based on the Crank-Nicolson method.

**Conclusions and Outlook**

Thus, we have reviewed the recent advances in quantum-mechanics based modelling of human perception and decision-making and presented several practical examples of quantum-mechanical models of optical illusions and opinion formation in social networks. We have also developed a computational code that the readers can use to create their own models. In the remainder of this section, we give several suggestions on how to efficiently use this computational code in different practical situations. We also outline several potential research directions, where the laws of quantum mechanics can be used to model human behaviour.

Generally speaking, the physical principles that underpin the operation of the computational code can be applied in many practical decision-making situations provided that the physical parameters of the model are chosen so that the code produces interpretable results. Knowing which parameters to select in order to obtain a physical meaningful result requires certain experience in numerical modelling (Sigrist, 2020). However, in general, the model will operate correctly if the reader uses common sense. In particular, it is important to remember that any model can produce plausible results only within the bounds of idealisations and assumptions made at the design stage. Consequently, when an attempt is made to apply an existing model to a new simulation scenario, it is advisable to verify whether the assumptions made in the existing model would be warranted in a different situation.

We also note that quantum-mechanics models of human behaviour are phenomenological. This means that, while such models describe the empirical relationship of phenomena to each other, they are not directly derived from the decision theory or theoretical and philosophical aspects of psychology. Furthermore, although there exists scientifically sound works suggesting that the human brain uses quantum computations (Kerskens & López Pérez, 2022), at the present those findings are not relevant to the theory of quantum cognition that uses the methods of quantum mechanics to explain human behaviour. Nevertheless, it is plausible that in the future both quantum-computational brain and quantum cognition approaches will be unified, revolutionising our understanding of how the functionality of the biological brain shapes human cognition and behaviour (Penrose, 1989).

The Schrödinger equation solved in the computational code is just one equation in the theory of quantum mechanics. For example, exploiting the physical property of spin that can be mathematically described as a vector that assumes not only the basis states $\uparrow$ and $\downarrow$ but also the states between them, it was demonstrated that quantum mechanics can be used to model paradoxical decision-making behaviours such as preference reversal (Maksymov & Pogrebna, 2023a). The key equation of that model is the Landau-Lifshitz equation. While that equation describes a purely quantum-mechanical phenomenon of magnetisation, it was derived phenomenologically using the methods of classical physics (Lakshmanan, 2011). Subsequently, models based on its equation can bridge the gap between the classical and quantum models of human cognition, enabling the researchers to solve a variety of complex social problems (Maksymov, 2024b).

**Data availability statement**

The computation code that accompanies this text and Supplementary Videos 1–3 can be accessed at https://github.com/IvanMaksymov/BehavDataSciCodes. The readers may also be interested in a relevant computational code that can be accessed at https://github.com/IvanMaksymov/OpinionPolarisation.

**References**

Abbas A. H., & Maksymov I. S. (2024). Reservoir computing using measurement-controlled quantum dynamics. *Electronics, 13(6)*, 1164.

Acemoglu, D., Como, G., Fagnani, F., & Ozdaglar, A. (2013). Opinion fluctuations and disagreement in social networks. *Math. Oper. Res., 38(1)*, 1–27.

Aerts, D., Sozzo, S., Tapia, J. (2012). *A Quantum Model for the Ellsberg and Machina Paradoxes*. In: Busemeyer, J.R., Dubois, F., Lambert-Mogiliansky, A., Melucci, M. (eds) *Quantum Interaction. QI 2012. Lecture Notes in Computer Science*, vol 7620. Berlin: Springer.

Aerts, D., Sozzo., S., & Tapia, J. (2014). Identifying quantum structures in the Ellsberg paradox. *Int. J. Theor. Phys., 53*, 3666–3682.

Aerts, D., & Arguëlles, J. A. (2022). Human perception as a phenomenon of quantization. *Entropy, 24*, 1207.

Aerts, D., & Beltran, L. (2022). A Planck radiation and quantization scheme for human cognition and language. *Front. Psychol., 13*, 850725.

Allan, B. B. (2018). Social Action in Quantum Social Science. *Millennium, 47*(1), 87–98.

Andrews, M. (1998). Wave packets bouncing off walls. *Am. J. Phys., 66*, 252–254.

Atmanspacher, H., & Filk, T. (2010). A proposed test of temporal nonlocality in bistable perception. *J. Math. Psychol., 54*, 314–321.

Bail, C. A., Argyle, L. P., Brown, T. W., Bumpus, J. P., Chen, H., Hunzaker, M. B. F., Lee, J., Mann, M., Merhout, F., & Volfovsky, A. (2018). Exposure to opposing views on social media can increase political polarization. *PNAS, 115(37)*, 9216–9221.

Basar-Eroglu, C., Mathes, B., Khalaidovski, K., Brand, A., & Schmiedt-Fehr, C. (2016) Altered alpha brain oscillations during multistable perception in schizophrenia. *Int. J. Psychophysiol., 103*, 118–128.

Bertlmann, R., & Friis., N. (2023). *Modern Quantum Theory*. Oxford: Oxford University Press.

Bompas, A., Campbell, A. E., & Sumner, P. (2020). Cognitive control and automatic interference in mind and brain: A unified model of saccadic inhibition and countermanding. *Psych. Rev., 127*(4), 524–561.

Busemeyer, J. R., & Bruza, P. D. (2012). *Quantum Models of Cognition and Decision*. Oxford: Oxford University Press.

Castellano, C., Fortunato, S., & Loreto, V. (2009). Statistical physics of social dynamics. *Rev. Mod. Phys., 81*, 591–646.

Castro, A. D. (2013). On the quantum principles of cognitive learning. *Behav. Brain Sci., 36*, 281–282.

Chen, X., Tsaparas, P., Lijffijt, J., & Bie, T. D. (2021). Opinion dynamics with backfire effect and biased assimilation. *PLoS ONE, 16*, e0256922.

Chen, R., Xiong, Y., Zhuge, S., Li, Z., Chen, Q., He, Z., Wu, D., Hou, F., & Zhou, J. (2023). Regulation and prediction of multistable perception alternation. *Chaos Solitons Fractals, 172*, 113564.

Clément, G., Allaway, H. C. M., Demel, M., Golemis, A., Kindrat, A. N., Melinyshyn, A. N., Merali, T., & Thirsk, R. (2015). Long-duration spaceflight increases depth ambiguity of reversible perspective figures. *PLOS ONE, 10*. https://doi.org/10.1371/journal.pone.0132317.

Dicker, G. (1993). *Descartes: An Analytical and Historical Introduction*. Oxford: Oxford University Press, Oxford.

Dittrich, T. (2014). 'The concept of information in physics': an interdisciplinary topical lecture. *Eur. J. Phys., 36*, 015010.

Dixon, G., Bullock, O., & Adams, D. (2019). Unintended effects of emphasizing the role of climate change in recent natural disasters. *Environ. Commun., 13*, 135–143.

Feynman, R., Leighton, R., & Sands, M. (1964). *The Feynman Lectures on Physics* (vol. 3). Pasadena: California Institute of Technology.

Fisher, G. H. (1967). Ambiguous figure treatments in the art of Salvador Dali. *Percept. Psychophys., 2*, 328–330.

Fleeson, W., & Wilt, J. (2010). The relevance of big five trait content in behavior to subjective authenticity: do high levels of within-person behavioral variability undermine or enable authenticity achievement? *J. Pers., 78*(4), 1353–1382.

Galam, S. (1997). Rational group decision making: A random field Ising model at T = 0. *Phys. A, 238*, 66–80.

Galam, S. (2005). Heterogeneous beliefs, segregation, and extremism in the making of public opinions. *Phys. Rev. E, 71*, 046123.

Galam, S. (2012). *Sociophysics: A Physicist's Modeling of Psycho-political Phenomena*. Berlin: Springer.

Galam, S., & Brooks, R. R. W. (2022). Radicalism: The asymmetric stances of radicals versus conventionals. *Phys. Rev. E, 105*, 044112.

Hu, H. (2017). Competing opinion diffusion on social networks. *R. Soc. Open Sci., 4*, 171160.

Inoue, M., & Nakamoto, K. (1994). Dynamics of cognitive interpretations of a Necker cube in a chaos neural network. *Prog. Theor. Phys., 92*, 501–508.

Interian, R., & Rodrigues, F. A. (2023). Group polarization, influence, and domination in online interaction networks: a case study of the 2022 Brazilian elections. *J. Phys. Complex., 4*, 035008.

Joos, E., Giersch, A., Hecker, L., Schipp, J., Heinrich, S. P., van Elst, L. T., & Kornmeier, J. (2020). Large EEG amplitude effects are highly similar across Necker cube, smiley, and abstract stimuli. *PLoS ONE, 15*, e0232928.

Kauffman, S. A., & Roli, A. (2022). What is consciousness? Artificial intelligence, real intelligence, quantum mind and qualia. *Biol. J. Linn. Soc., 139*, 530–538.

Kerskens, C. M, & López Pérez, D. (2022). Experimental indications of non-classical brain functions. *J. Phys. Commun., 6*, 105001.

Khalil, E. L. (2021). Why does Rubin's vase differ radically from optical illusions? Framing effects contra cognitive illusions. *Front. Psychol., 12*, 597758.

Khrennikov, A. (2006). Quantum-like brain: "Interference of minds". *Bio Systems, 84*(3), 225–241.

Kim, P. (2017). MATLAB Deep Learning With Machine Learning, Neural Networks and Artificial Intelligence. Berkeley, CA: Apress.

Korn, H., & Faure, P. (2003). Is there chaos in the brain? II. Experimental evidence and related models. *C. R. Biol., 326*, 787–840

Kornmeier, J., & Bach, M. (2005) The Necker cube–an ambiguous figure disambiguated in early visual processing. *Vision Res., 45*, 955–960.

Kragh. H. (2000). Max Planck: the reluctant revolutionary. *Phys. World*. https://physicsworld.com/a/max-planck-the-reluctant-revolutionary/.

Kumar, V., & Pravinkumar, P. (2023). Quantum random number generator on IBM QX. *J. Cryptogr. Eng.*, https://doi.org/10.1007/s13389-023-00341-1.

Lakshmanan M., (2011). The fascinating world of the Landau–Lifshitz–Gilbert equation: an overview. *Phil. Trans. R. Soc. A., 369,* 1280–1300.

Larsen, S. (2022). Double Slit Simulation – Schrodinger FDTD. Retrieved 28 March 2024, from https://blog.c0nrad.io/posts/sim-9-double-slit-schrodinger-fdtd/.

Liebertz, S., & Bunch, J. (2021). Backfiring frames: abortion politics, religion, and attitude resistance. *Politics Relig., 14*, 403–430

Lindstrøm, T .C., & Kristoffersen, S. (2001). 'Figure it out!' Psychological perspectives on perception of migration period animal art. *Nor. Archaeol. Rev., 34*, 65–84.

Lo, C., & Dinov, I. (2011). Investigation of optical illusions on the aspects of gender and age. *UCLA USJ, 24*, 5–10.

Long, G. M., & Toppino, T. C. (2004). Enduring interest in perceptual ambiguity: alternating views of reversible figures. *Psychol. Bull., 130*, 748–768.

Magnuski, H. (2023). Neckerworld – A Computer Vision Game. Retrieved 28 March 2024, from http://http://www.neckerworld.com/.

Maksymov, I. S., & Pogrebna, G. (2023). Linking Physics and Psychology of Bistable Perception Using an Eye Blink Inspired Quantum Harmonic Oscillator Model. arXiv preprint arXiv:2307.08758.

Maksymov, I. S., & Pogrebna, G. (2023). The physics of preference: Unravelling imprecision of human preferences through magnetisation dynamics. arXiv preprint arXiv:2310.00267.

Maksymov, I. S., & Pogrebna, G. (2024). Quantum-Mechanical Modelling of Asymmetric Opinion Polarisation in Social Networks. *Information, 15*, 170.

Maksymov, I. S. (2024). Quantum-inspired neural network model of optical illusions. *Algorithms, 17(1)*, 30.

Maksymov, I. S. (2024). Physical Reservoir Computing Enabled by Solitary Waves and Biologically Inspired Nonlinear Transformation of Input Data. *Dynamics, 4(1)*, 119–134.

Maksymov, I. S. (2024). Magnetism-inspired quantum-mechanical model of gender fluidity. *Psychol. J. Res. Open, 6(1)*, 1–7. DOI: 10.31038/PSYJ.2024614.

Martínez-Martínez, I., & Sánchez-Burillo, E. (2016). Quantum stochastic walks on networks for decision-making. *Sci. Reps., 6*, 23812.

McKenna, T .M., McMullen, T. A., & Shlesinger, M. F. (1994). The brain as a dynamic physical system. *Neuroscience, 60*, 587–605.

McManus, I. C., Freegard, M., Moore, J., & Rawles, R. (2010). Science in the Making: Right Hand, Left Hand. II: The duck-rabbit figure. *Laterality, 15(1-2)*, 166–185.

Meilikhov, E. Z., & Farzetdinova, R. M. (2019). Bistable perception of ambiguous images: simple Arrhenius model. *Cogn. Neurodyn., 13*, 613–62.

Mena, A. (2023). Solving the 2D Schrödinger equation using the Crank-Nicolson method. Retrieved 28 March 2024, from https://artmenlope.github.io/solving-the-2d-schrodinger-equation-using-the-crank-nicolson-method/.

Mindell, A. (2012). *Quantum mind: The edge between physics and psychology*. Florence: Deep Democracy Exchange.

Moreira, C., Tiwari, P., Pandey, H. M., Bruza, P., & Wichert, A. (2020). Quantum-like influence diagrams for decision-making. *Neural Netw., 132*, 190–210.

Necker, L. A. (1832). Observations on some remarkable optical phenomena seen in Switzerland; and on an optical phenomenon which occurs on viewing a figure of a crystal or geometrical solid. *Lond. Edinb. Philos. Mag. J. Sci., 1*, 329–337.

Nicholls, M. E. R., Churches, O., & Loetscher, T. (2018). Perception of an ambiguous figure is affected by own-age social biases. *Sci. Reps., 8(1)*, 12661.

Nielsen, M., & Chuang, I. (2002). *Quantum computation and quantum information*. Oxford: Oxford University Press.

Nietzsche, F. (1886). *Beyond Good and Evil* (translated by Helen Zimmern, 1906; reprinted in 1997). New York: Courier Dover Publications.

Nyhan, B., & Reifler, J. (2015). Does correcting myths about the flu vaccine work? An experimental evaluation of the effects of corrective information. *Vaccine, 33(3)*, 459–464.

Ortega y Gasset, J. (1966). *Obras Completas, Vol. I* (p. 322). Madrid: Revista de Occidente.

Parkkonen, L., Andersson, J., Hämäläinen, M., & Hari, R. (2008). Early visual brain areas reflect the percept of an ambiguous scene. *PNAS, 105(51)*, 20500–20504.

Patry, J.-L. (2011). Methodological consequences of situation specificity: biases in assessments. *Front. Psychol., 2*, 18. 10.3389/fpsyg.2011.00018.

Penrose, Roger (1989). *The Emperor's New Mind*. New York, New York: Penguin Books.

Piantoni, G., Romeijn, N., Gomez-Herrero, G., Van der Werf, Y. D., & Van Someren, E. J. W. (2017). Alpha power predicts persistence of bistable perception. *Sci. Rep., 7*, 5208.

Pothos, E. M., & Busemeyer, J. R. (2022). Quantum cognition. *Annu. Rev. Psychol., 73*, 749–778.

Qureshi, A. W., Monk, R. L., Samson, D., & Apperly, I. A. (2020). Does interference between self and other perspectives in theory of mind tasks reflect a common underlying process? Evidence from individual differences in theory of mind and inhibitory control. *Psychon. Bull. Rev., 27(1)*, 178–190.

Redner, S. (2019). Reality-inspired voter models: A mini-review. *C. R. Phys.*, 20, 275–292.

Sigrist, J. (2020). *Numerical Simulation, An Art of Prediction 1: Theory*. United Kingdom: Wiley.

Shapiro, A. G., & Todorović, D. (2017). *The Oxford Compendium of Visual Illusions*. Oxford: Oxford University Press.

Shimaoka, D., Kitajo, K., Kaneko, K., & Yamaguchi, Y. (2010). Transient process of cortical activity during Necker cube perception: from local clusters to global synchrony. *Nonlinear Biomed. Phys., 4*, S7.

Schilpp, P. A. (1949). *Albert Einstein: Philosopher-Scientist*. Evanston: The Library of Living Philosophers, Inc.

Smith, R. (2020). *Quantum Physics For Beginners*. Seattle: Amazon Publishing.

Stonkute, S., Braun, J., & Pastukhov, A. (2012). The role of attention in ambiguous reversals of structure-from-motion. *PLoS ONE, 7*, e37734.

Swire-Thompson, B., DeGutis, J., & Lazer, D. (2020). Searching for the Backfire Effect: Measurement and Design Considerations. *J. Appl. Res. Mem. Cogn.*, 9(3), 286–299.

Symul, T., Assad, S. M., & Lam, P. K. (2011). Real time demonstration of high bitrate quantum random number generation with coherent laser light. *Appl. Phys. Lett., 98*, 231103.

Sznajd-Weron, K., & Sznajd, J. (2000). Opinion evolution in closed community. *Int. J. Mod. Phys. C*, 11, 1157–1165.

Sullivan, D. M. (2000). *Electromagnetic Simulation Using the FDTD Method*. New York: IEEE Press.

Takeno, J. (2013). *Creation of a Conscious Robot: Mirror Image Cognition and Self-Awareness*. Boca Raton: CRC Press.

Tesař, J. (2015). Quantum Theory of International Relations: Approaches and Possible Gains. *Human Affairs, 25*(4), 486-502.

Thomas, J. W. (1995). *Numerical Partial Differential Equations: Finite Difference Methods. Texts in Applied Mathematics*. Vol. 22. Berlin: Springer-Verlag.

Troje, N. F., & McAdam, M. (2010). The viewing-from-above bias and the silhouette illusion. *i-Perception, 1(3)*, 143–148.

Walker, J. S. (2021). *Physics*. 5th ed. London: Pearson.

Wallis, G., & Ringelhan, S. (2013). The dynamics of perceptual rivalry in bistable and tristable perception. *J. Vis., 13(2)*, 24.

Wei, M., al Nowaihi, A., & Dhami, S. (2019). Quantum decision theory, bounded rationality and the Ellsberg Paradox, *Stud. Microecon., 7*, 110–139.

Wendt, A. (2015). *Quantum Mind and Social Science*. Cambridge: Cambridge University Press.

Wang, X., Sang, N., Hao, L., Zhang, Y., Bi, T., & Qiu, J. (2017). Category selectivity of human visual cortex in perception of Rubin face-vase illusion. *Front. Psychol., 8*, 1543.

Wang, P.-Y., Xu, C-H., Wang, P.-Y., Huang, H-Y., Chang, Y-W., Cheng, J-H., Lin, Y.-H., & Cheng, L.-P. (2021). Game Illusionization: A Workflow for Applying Optical Illusions to Video Games. In *The 34th Annual ACM Symposium on User Interface Software and Technology (UIST '21). Association for Computing Machinery*. New York, NY, USA, 1326–1344. https://doi.org/10.1145/3472749.3474824.

Ward, E. J., & Scholl, B. J. (2015). Stochastic or systematic? Seemingly random perceptual switching in bistable events triggered by transient unconscious cues. *J. Exp. Psychol. Hum. Percept. Perform., 41(4)*, 929–939.

Washburn, M., Reagan, C., & Thurston, E. (1934). The comparative controllability of the fluctuations of simple and complex ambiguous perspective figures. *Am. J. Psychol., 46*, 636–638.

Wilson, M., Hecker, L., Joos, E., Aertsen, A., Tebartz van Elst, L., & Kornmeier, J. (2023). Spontaneous Necker-cube reversals may not be that spontaneous. *Front. Hum. Neurosci., 17*, 1179081.

Yamamoto, S., & Yamamoto, M. (2006). Effects of the gravitational vertical on the visual perception of reversible figures. *Neurosci. Res., 55*, 218–121

Zurek, W. H. (2003). Decoherence, einselection, and the quantum origins of the classical. *Rev. Mod. Phys., 75*, 715–775.

Zurek, W. H. (2009). Quantum Darwinism. *Nat. Phys., 5*, 181–188.

Zurek, W. H. (2015). Classical selection and quantum Darwinism. *Physics Today, 68*(5), 9–10.